\documentclass{kluwer}
\usepackage{epsfig}

\newcommand{\citen}[1]{\citeauthor{#1},\ \citeyear{#1}}
\newcommand{\citet}[1]{\citeauthor{#1}\ (\citeyear{#1})}

\newcommand{\apj}{{\it Astrophys.~J.}}			
\newcommand{\apjl}{{\it ApJ}}

\newcommand{\aap}{{\it Astron.~Astrophys.}}

\newcommand{\pasj}{{\it PASJ}}

\newcommand{\solphys}{{\it Solar~Phys.}}

\newcommand{\nat}{{\it Nature}}

\newcommand{\vc}[1]{\mbox{\boldmath $#1$}}

\newcommand{\dv}[2]{\frac{{\rm d}#1}{{\rm d}#2}}
\newcommand{\ov}{\overline}
\newcommand{\scri}{\scriptsize} 

\begin{document}
\begin{article}
\begin{opening}

\title{The Effect of Abnormal Granulation on Acoustic Wave Travel Times and
Mode Frequencies}

\author{K. \surname{Petrovay}$^{1,2}$%\email{K.Petrovay@astro.elte.hu}
}

\author{R. \surname{Erd\'elyi}$^{1}$}
\author{M. J. \surname{Thompson}$^{1}$}
\institute{$^{1}$Solar Physics and Upper Atmosphere Research Group, 
Department of Applied Mathematics,
University of Sheffield, Hicks Building, Hounsfield Road. Sheffield S3 7RH,
U.K.\\
$^{2}$E\"otv\"os University, Department of Astronomy, Budapest, Pf. 32, 
H-1518 Hungary}
%\email{K.Petrovay@astro.elte.hu; Robertus@sheffield.ac.uk; Michael.Thompson@sheffield.ac.uk}

\date{Received;accepted}

\runningtitle{Effect of Abnormal Granulation on Sound Waves}
\runningauthor{Petrovay et al.}

\begin{abstract}
Observations indicate that in plage areas (i.e. in active regions outside
sunspots) acoustic waves travel faster than in quiet sun, leading to shortened
travel times and higher $p$-mode frequencies. Coupled with the 11-year
variation of solar activity, this may also explain the solar cycle variation of
oscillation frequencies. While it is clear that the ultimate cause of any
difference between quiet sun and plage is the presence of magnetic fields of
order 100 G in the latter, the mechanism by which the magnetic field exerts its
influence has not yet been conclusively identified. One possible such mechanism
is suggested by the observation that granular motions in plage areas tend to be
slightly ``abnormal'', dampened compared to quiet sun.

In this paper we consider the effect that abnormal granulation observed in
active regions should have on the propagation of acoustic waves. Any such
effect is found to be limited to a shallow surface layer where sound waves
propagate nearly vertically. The magnetically suppressed turbulence implies
higher sound speeds, leading to shorter travel times. This time shift
$\Delta\tau$ is independent of the travel distance, while it shows a
characteristic dependence on the assumed plage field strength. As a consequence
of the variation of the acoustic cutoff with height, $\Delta\tau$ is expected
to be significantly higher for higher frequency waves within the observed
regime of 3--5 mHz. The lower group velocity near the upper reflection point
further leads to an increased envelope time shift, as compared to the phase
shift. $p$-mode frequencies in plage areas are increased by a  corresponding
amount, $\Delta\nu/\nu=\nu\Delta\tau$. These characteristics of the time and
frequency shifts are in accordance with observations. The calculated overall
amplitude of the time and frequency shifts are comparable to, but still
significantly (factor of 2 to 5) less than suggested by measurements. 
\end{abstract}


\end{opening}

\section{Introduction}

Early observations proved that the global acoustic frequencies show small but
significant  and systematic correlations with the solar cycle. Although the
acoustic modes are strongly evanescent in the atmosphere, changes of magnetic
fields and mean temperature at the  boundary layer (i.e. lower atmosphere)
could play a rather important role in the determination  of their frequencies
acting as boundary conditions in an eigenvalue problem when determining these 
frequencies, since magnetic or flow fields in the boundary layer change the
mean elasticity of  the boundary itself or alter the upper turning points.
These effects contribute to small corrections to the eigenfrequencies. The
pioneering papers by \citet{Campbell+Roberts} and \citet{Roberts+Campbell},
opened a new series of studies of the coupling of solar global oscillations to
the lower solar atmosphere. The influence of an atmospheric coherent magnetic 
field on $p$- and $f$-mode frequencies was evaluated theoretically for a simple
and elegant model. It was shown that, at low to moderate degree $l$, an
increase in chromospheric magnetic field leads to a frequency {\it increase}
for the $n=1$ $p$-mode, whereas the overtones ($n$=2, 3,  etc.) suffer a
frequency decrease. It was shown that at high $l$, all the $p$-modes suffer a
frequency  decrease. The effect of the coherent magnetic canopy on the solar
acoustic modes was extended to  allow for variations in height of the magnetic
canopy (\citen{Evans+Roberts1}, \citeyear{Evans+Roberts2}). It was found that 
changes in chromospheric magnetism can be manifested in $p$- and $f$-mode data
sets gathered at different phases of the solar cycle. These predictions of
solar-cycle variability in high-degree $p$-mode frequencies from a simple
model of the magnetic canopy which permeates the solar atmosphere were
compared with the observations of \citet{Libbrecht+Woodard}. Good agreement
was  found with the observed frequency shifts for modes of frequency less than
4 mHz, through a mechanism  in which an increasing magnetic field induces
``stiffening'' of the Sun's lower atmosphere.

The influence of short-scale motions (i.e. granulation) modelled as a random
flow on $f$-mode frequencies  was first evaluated by 
\citeauthor{Murawski+Roberts2} (\citeyear{Murawski+Roberts1},b; see also 
\citen{Murawski+Goossens}, \citen{Ghosh+}, \citen{Gruzinov:random.waves},  and
\citen{Medrek+}). \citeauthor{Erdelyi+Kerekes+Mole1}
(\citeyear{Erdelyi+Kerekes+Mole1}, \citeyear{Erdelyi+Kerekes+Mole2}) have
revisited this problem.  The $f$-mode is essentially a surface wave; hence the
mode frequencies are less likely to be influenced  by the solar stratification.
Most probably the discrepancies are the result of near surface  mechanisms,
such as interactions with surface or sub-surface magnetic fields and flows.
\citeauthor{Erdelyi+Kerekes+Mole1} followed the general approach of
\citeauthor{Murawski+Roberts1}, which is a valuable one, but corrected certain
errors which appeared in that paper. The simple model used by
\citeauthor{Murawski+Roberts1} and \citeauthor{Erdelyi+Kerekes+Mole1}  gives a
deviation of the $f$-modes from the theoretically predicted parabolic ridges
which agrees  qualitatively with observations. They found that turbulent
background flows can {\it reduce} the  eigenfrequencies of global solar
$f$-modes by several percents, as found in observations at high spherical
degree. Extensive numerical simulations of the outer parts of the Sun carried
out by, e.g.,  \citet{Rosenthal+} demonstrated and quantified the influence of
turbulent convection on solar  oscillation frequencies.

In the lower part of the solar atmosphere,  i.e. in the {\it boundary layer}
between the  solar interior and the magnetically dominant solar corona  there
are both coherent and  random components of the velocity and magnetic fields,
each of which may contribute to  the frequency shifts and line widths of the
global solar acoustic oscillations. Before  a comprehensive model including all
these effects is constructed, it is vital for a  better understanding to
estimate the importance of these various effects. Random flows  (e.g. turbulent
granular motion), coherent flows (meridional flows or the near-surface 
component of the differential rotation), random magnetic fields (e.g. the
continuously  emerging tiny magnetic fluxes or magnetic carpet) and coherent
fields (large loops and  their magnetic canopy region) may each  affect the
acoustic perturbations.  Some of these effects may be more important than
others. The magnitude of these corrections has to be estimated one by one and
it is suspected that, unfortunately, they  all may contribute to line widths or
frequency shifts of the global acoustic oscillations on a rather equal basis.
In what follows we recall briefly some previous findings, in order to have a
basis for  comparison with the present results.

The magnetic field in the photosphere has, like the granulation, a random
component. High resolution magnetograms reveal that outside active regions the
solar surface  is covered with a mixed polarity network, which has been termed
the {\it magnetic carpet} (\citen{Title+Schrijver:carpet}).
\citet{Erdelyi+Kerekes+Mole2} investigated the influence of this disorganised,
small-scale lower atmospheric field on the $f$-mode frequencies. The magnetic
carpet was modelled as a time-independent,  stochastic field. Since, depending
on their spherical degree, some $f$-modes may have a life-time  comparable to
the characteristic replacement time (of the order of tens of hours) of the
magnetic  carpet, this limits the validity of the Erd\'elyi et al. study. 
Nonetheless they found that a time-independent random magnetic  field can
significantly {\it increase} the $f$-mode frequencies. 

Flow fields at the lower atmospheric boundary layer may also be random or
coherent. By inverting the  observational data of solar global oscillations one
could potentially reconstruct the global flow  structures. Large-scale
sub-surface flows were found by this technique (e.g. \citen{Braun+Fan}). To the
best of our knowledge, \citet{Erdelyi+Varga+Zetenyi} were the first who studied
the effect of a  sub-surface motion on magnetoacoustic-gravity surface
waves, representing the $f$-mode in a model  of the solar interior - solar
atmosphere interface. They found that the flow causes a shift of the  forward
and backward propagating magnetoacoustic-gravity modes, which in certain cases bifurcate. Erd\'elyi
\& Taroyan  (\citeyear{Erdelyi+Taroyan1999}, \citeyear{Erdelyi+Taroyan2001})
generalised the model by  allowing the temperature to increase linearly  with
depth in the sub-surface  zone. They derived the dispersion relation and
analytical formulae for  the frequencies of $p$- and $f$-modes in the  limit of
small wave numbers. Numerical  solutions were  presented for other cases.   

All these global studies suggest that the strong magnetic perturbations
associated with active regions can have an important effect on the propagation
and reflection of sound waves near the solar surface. Time-distance and other
types of local helioseismic studies confirm that the  travel time of sound
waves is significantly different in the subphotospheric layers of active
regions than in the quiet Sun. In plage areas, outside sunpots, the travel time
is generally found to be shorter than in quiet sun areas (e.g.
\citen{Chou:acoustic.imaging}; \citen{Kosovichev:tomography};
\citen{Hughes+:2AR}). $p$-mode frequencies are found to be systematically
higher in plage areas than in quiet sun (\citen{Rajaguru+:ringdiag}). All this
implies either a different form of the dispersion relation in plage areas, or a
perturbed background stratification (and consequently a perturbed sound speed
and acoustic cutoff).

Several possible physical mechanisms have been proposed to explain such 
discrepancies in magnetized regions. \citet{Zweibel+Bogdan:sound.fibrils}
considered the effect that the presence of an ensemble of magnetic flux fibrils
in plage regions has on sound wave propagation. \citet{Bruggen+Spruit}, in
turn, discussed the effects an altered temperature profile could have on sound
wave travel times. One plausible effect of plage magnetic fields has, however,
apparently not been considered from the point of view of acoustic propagation:
the magnetic suppression of turbulent motions, observable in the form of
``abnormal granulation'' (\citen{Dunn+Zirker:filigree}). The aim of the present
work is to complement previous work by considering the potential effect of
abnormal granulation on sound wave travel times in plage areas. Note that here
we only consider the effect of turbulence on the propagation of coherent waves,
disregarding wave excitation and scattering.

Our method of calculation, based on the standard ray approximation and on the
assumption of isotropic turbulence, is described in Section 2. Results are
presented in Section 3, while Section 4 concludes the paper.

\section{Method}

\subsection{Ray approximation}

The ray approximation, a widely used method of time-distance helioseismology,
is valid in the short wavelength limit $\lambda\ll H$, where $H$ is the density
scale height. As in the solar convective zone the integral scale of
turbulence is $l\ge H$, it is then consistent to use ray approximation also for
the study of wave-turbulence interactions.

The local dispersion relation for a travelling wave of wave vector $\vc k$ in
the stratified superadiabatic fluid reads 
\begin{equation} \omega^2=\omega_c^2 + c^2 k^2  , 
\label{eq:disprel0}   
\end{equation} 
where $c$ is the adiabatic sound speed, and $\omega_c \simeq c/2H $ is the
acoustic cutoff frequency. Our assumption $\lambda\ll H$ would clearly imply
that the wave frequency is high compared to the acoustic cutoff:  $\omega\sim
c/\lambda\gg c/H\sim \omega_c$. On the other hand, the upper reflection of the
sound waves, responsible for forming the wave duct, is due to the sudden
increase of the acoustic cutoff in the photosphere. Thus, the ray approximation
is bound to break down in a shallow layer below the surface: its widespread use
is based on the circumstance that near the surface the waves propagate nearly
vertically, so their propagation is not expected to be strongly  influenced by
interference with scattered or reflected waves. In line with this reasoning,
the neglect of $\omega_c$ in the dispersion relation is consistent with the use
of the ray approximation. Sound waves then become nondispersive, their upper
reflection imposed externally at a prescribed depth $z_t$ where
$\omega_c=\omega$. (Indeed, \citen{dSilva:sspot} finds that acoustic wave
packets are only very slightly dispersive.) The effect of turbulence on
nondispersive waves can then be described by a correction factor $F_t$:
\begin{equation} \omega=F_t c k .  \label{eq:disprel1}   
\end{equation} In the
general case $F_t$ is anisotropic; however, for the present calculation we
consider isotropic turbulence, resulting in a scalar $F_t$.  The form of $F_t$
will be discussed in Section~2.2 below. Turbulent convection is certainly
anisotropic and inhomogeneous; however, presently very little is known about
how this anisotropy differs between abnormal granulation and normal
granulation. Thus, for this first exploratory study we consider it more
appropriate to limit the number of unknown parameters by restricting attention
to isotropic turbulence.

As the magnetic suppression of turbulence is expected to be confined to rather
shallow depths, we use a Cartesian setup. Then the ray equations forming the
basis of time-distance helioseismology (\citen{dSilva:foundations}) reduce to
very simple forms. Introducing the effective sound speed $\vc C=F_t c\,\vc k/k$, 
the equations can be recast as a relation for the ray path:
\begin{equation} \dv zx = \frac{C_z}{C_x}  ,
\end{equation}
or in integral form
\begin{equation} x=\int \frac{C_x}{C_z} \,{\rm d}z  ,  \label{eq:raypath0} 
\end{equation}
and an integral expression (taken along the ray path) for the acoustic travel
time of a wave packet 
\begin{equation} \tau=\int \frac{{\rm d}s}C .   \label{eq:travtime0} 
\end{equation}
Substituting equation (\ref{eq:disprel1}) into equations 
(\ref{eq:raypath0})--(\ref{eq:travtime0}) yields the travel time--distance
relation in a parametric form.

\subsection{Magnetic suppression of turbulence}

Throughout this paper we assume that the turbulent motions are very subsonic:
$\beta\equiv v/c\ll 1$. This is a good approximation in subphotospheric layers
of the Sun. In this limit Taylor's ``frozen turbulence'' hypothesis
(\citen{Taylor:frozen}) applies, i.e. the turbulent flow can be regarded as
random in space but stationary in time (over the timescales relevant for sound
propagation through the turbulent fluctuations).

While the random fluctuations also lead to a random refraction of the ray path,
in homogeneous and isotropic turbulence the {\it mean} wave path remains
straight; thus, the correction factor $F_t$ in equation (\ref{eq:disprel1}) is
a scalar, and the mean ray propagation is only affected by the statistics of
the flow component $v=\vc v\cdot\vc k/k$ parallel to the wave vector; the problem is
thus reduced to one dimension.

The effect of homogeneous and isotropic turbulence on wave propagation in the
ray approximation is then twofold. Scalar fluctuations in the thermodynamical
variables imply fluctuations in the sound speed $c$, whereas random flows
introduce Doppler corrections to the wavenumber. Ensemble averaging over
different realizations of the turbulent flow, neither of these effects will
cancel out, as already pointed out by \citet{Brown:turb.oscill}. The reason is
simple to understand: in the frozen turbulence limit the wave packet will spend
more time in those parts of the flow where the sound speed is below average, or
$v <0$. (Note that in the opposite limit $\beta\gg 1$, in contrast, any net
mean effect of turbulence should cancel out in the ray approximation.)

Let us consider the first of these effects: a straight acoustic ray propagating
in a medium with random flows, of PDF\footnote{The probability distribution
function (PDF) is here meant in the sense of the frequency interpretation of
probability, i.e. $f(v)\,dv$ is the fraction of time when the velocity has a
value between $v$ and $v+dv$. Consequently, averages mean temporal averages
throughout this paper.} $f(v)$, aligned with the direction of propagation. The
mean travel time over unit distance is then clearly
\begin{equation} t=\int_{-\infty}^\infty \frac{f(v)\,{\rm d}v}{c+v},   
\end{equation}
so the turbulence correction factor due to the Doppler effect is
\begin{equation} F_{t,D}=\left[ \int_{-\infty}^\infty \frac{f(\beta)\,{\rm
  d}\beta}{1+\beta}\right]^{-1}  \simeq 1-\ov{\beta^2} ,    \label{eq:beta}
\end{equation}
where the second approximate equality follows from an expansion in $\beta$.

Analoguously, the sound speed fluctuations $c'$ give rise to a correction factor
\begin{equation} F_{t,c}\simeq 1-\ov{\delta^2} ,
\end{equation}
with $\delta\equiv c'/c$. In homogeneous isotropic turbulence scalar and vector
fluctutations are uncorrelated by assumption,\footnote{
The correlation of a fluctuating scalar field $\alpha$ and a fluctuating
vector field $\vc u$ is a mean vector field $\langle\alpha\vc u\rangle$. If
non-vanishing, this field selects a preferred direction in space, contradicting
the assumption of isotropy.}
so the two effects act
independently, leading to a total turbulent correction of
\begin{equation} F_t=1-(\ov{\beta^2}+\ov{\delta^2}) . 
\end{equation}
In what follows, the overbars will be omitted, denoting by $\beta$ and $\delta$
the r.m.s. values.

\begin{figure}
\centerline{
\includegraphics*[width=\columnwidth]{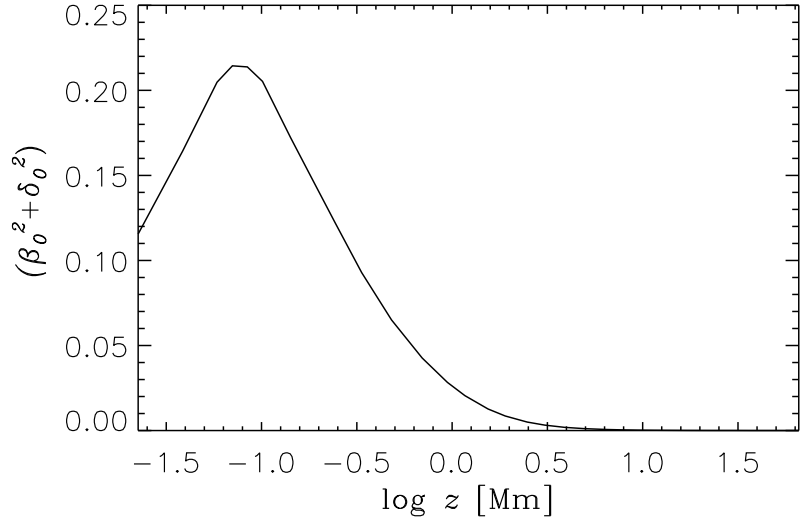}
}
\caption{Mean square velocity fluctuations ($\beta^2$) and sound speed
fluctuations ($\delta^2$), normalized by $c^2$, as a function of depth in the
solar convective zone.  $z=0$ corresponds to a continuum optical depth of unity.
The index '0' on the ordinate refers to quiet sun values.}
\label{fig:turb_z}
\end{figure}

The distribution of $({\beta^2}+{\delta^2})$ in the solar convective zone is
shown in Figure \ref{fig:turb_z}.  For this and all other figures, the mean
thermal stratification of the solar envelope was taken from the model of
\citet{Guenther}, while for the turbulent velocities, the model of \citet{UKX}
was used. It is clear that the effect of turbulence on sound propagation is
limited to the uppermost few megameters. 

\begin{figure}
\centerline{
\includegraphics*[width=\columnwidth]{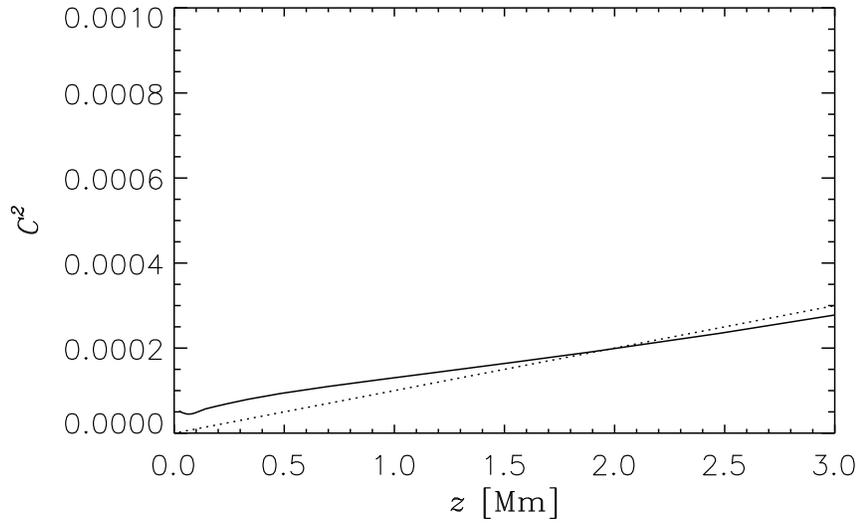}
}
\caption{Square of the effective sound speed $C=F_{t,0} c$, including the
turbulent reduction, as a function of depth in the solar convective zone.
(Units are megameters for depth and Mm/s for the sound speed.)}
\label{fig:cs2_z}
\end{figure}

The square of the effective sound speed $C=F_{t,0} c$, including the turbulent
reduction, is displayed in Figure~\ref{fig:cs2_z}. A linear fit of the form  
\begin{equation} C^2=mz \qquad m=10^{-4}         \label{eq:csappr}     
\end{equation} 
(dashed line) is found to be a
tolerable representation that will greatly simplify the calculation and make an
analytical solution possible. Such a linear dependence of the squared sound
speed on depth is also characteristic of polytropic atmospheres  with uniform
gravitational acceleration. Here, however, we do not rely on any other
assumption than equation~(\ref{eq:csappr}). (Indeed, using a polytrope with the
above value for $m$ would lead to completely inappropriate acoustic cutoff
frequencies and upper reflection depths.)

In plage areas,  magnetic fields will tend to suppress turbulent motions. On
energetic grounds one expects this reduction to have the form
\begin{equation} R\equiv 
   \frac{{\beta_1^2}+{\delta_1^2}}{{\beta_0^2}+{\delta_0^2}}=
   \frac 1{ 1+(B/B_e)^2}  ,              \label{eq:R}
\end{equation}
where the subscripts 1 and 0 refer to plage and quiet sun values, respectively;
$B$ is the magnetic flux density, $\rho$ is the mass density, and
\begin{equation} B_e=(2\pi\rho\ov{v^2})^{1/2}  
\end{equation}
is the turbulent equipartition field strength.

The reasoning behind equation (\ref{eq:R}) goes as follows. The equipartition
field $B_e$ is the field strength where the magnetic energy density equals the
kinetic energy density of turbulent motions. Then for $B\gg B_e$ the strong
field will effectively inhibit overturning turbulent motions $(R\rightarrow
0)$, while for $B\ll B_e$ we expect that the field cannot significantly damp
turbulence $(R\rightarrow 1)$, as its energy is not high enough to compensate
the work done by the buoyancy forces driving the turbulent convection. Thus, we
expect that the energy of the turbulence will be reduced by a factor $1/R$ of
order unity around $B_e$, i.e. $1/R\sim E_{\mbox{\scri turb}} / (E_{\mbox{\scri
turb}} + E_{\mbox{\scri mag}})$, as expressed in equation (\ref{eq:R}).

Thus, in a plage area the dispersion relation can be written as
\begin{equation} \omega=FC k ,
\end{equation}
where $C=F_{t}c $ is the quiet-sun sound speed (including turbulence effects),
while
\begin{equation} F=1+(1-R)({\beta_0^2}+{\delta_0^2})  \end{equation}
is the correction factor due to the magnetic field. Clearly $F>1$: the
suppression of turbulence in active regions leads to less turbulent
reduction of the sound speed there, i.e. sound waves will 
tend to propagate faster in
plage than in quiet sun due to this effect.

\begin{figure}
\centerline{
\includegraphics*[width=\columnwidth]{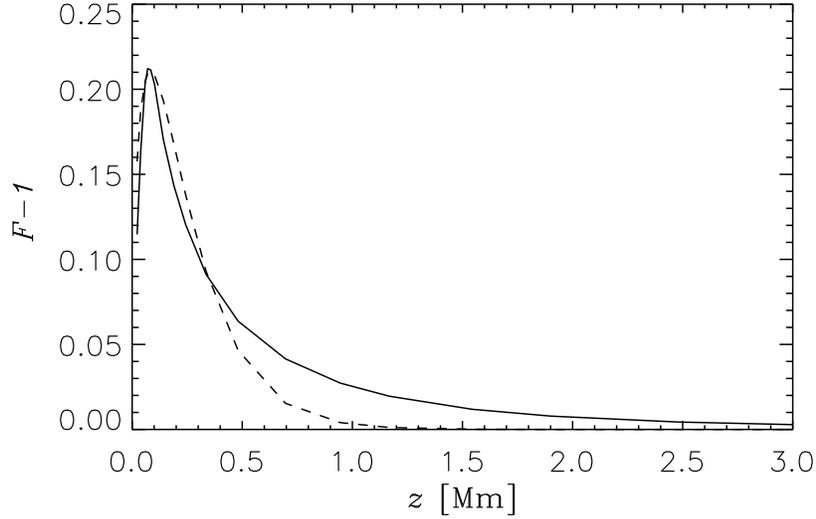}
}
\caption{Magnetic correction to the sound speed in a plage area of field
strength 400 G {\it (solid line),} and the analytic fit given in equation 
(\ref{eq:Fappr})}
\label{fig:F_z}
\end{figure}

Figure \ref{fig:F_z} shows $(F-1)$ as a function of depth for 
a presumed uniform field strength $B=400\,$G.  It
is apparent that significant magnetic reduction of turbulence is limited to
the uppermost 1 Mm approximately. This shallow layer of magnetically suppressed
turbulence is observed in the form of abnormal granulation. We find that
$F(B,z)$ can be roughly approximated by the simple analytic function
\begin{equation} F=1+f_1 \,\frac{(B/B_e)^2}{1+(B/B_e)^2} e^{-\alpha z}(z/2)^{1/2} 
   \label{eq:Fappr} \qquad f_1=1.8, \alpha=6 ,
\end{equation}
with $z$ in megameters (dashed line on Fig.~\ref{fig:F_z}). 

\section{Results}

Next, we substitute equations (\ref{eq:csappr})--(\ref{eq:Fappr}) into the ray
equations (\ref{eq:raypath0}) and (\ref{eq:travtime0}). For the ray path we
find that, for quiet sun, for fixed horizontal wave number $k_x$ and frequency
$\omega$, the lower reflection depth $a$, the travel distance $d$, and the
travel time $\tau$ are given by
\begin{equation}
a=\omega^2/mk_x^2 \qquad d=\pi a \qquad \tau=\pi (m/a)^{1/2}   ,
     \label{eq:wavepath}
\end{equation}
(see e.g. \citen{Bruggen+Spruit} for details). 

As the magnetic perturbation only extends down to a depth of $z_0\sim
1\,$Mm$\ll a$, for the calculation of the perturbations in travel distance and
travel time the integrals only need to be evaluated down to $z_0$. Expanding
the integrands to first order in $(F-1)$, and, upon substitution of
equation~(\ref{eq:Fappr}), also in $z/a$ results in
\begin{equation} 
   \Delta d\equiv d_{\mbox{\scri\rm plage}}
   -d_{\mbox{\scri\rm quiet sun}} =
    -\, f_1 \alpha^{-2}(2a)^{-1/2} 
   \left[ {e^{-\alpha\,z_0}}(1+\alpha\,z_0) -1\right] ,
\end{equation}
%\begin{eqnarray} && \Delta d\equiv d_{\mbox{\scri\rm plage}}
%     -d_{\mbox{\scri\rm quiet sun}} =
%  \nonumber \\
%   && -\, f_1 \alpha^{-2}(2a)^{-1/2} 
%   \left[ {e^{-\alpha\,z_0}}(1+\alpha\,z_0) -1\right] 
%\end{eqnarray}
with $d$ and $a$ in megameters. Evaluation of this formula shows that for
$a>3\,$Mm, $\Delta d<10\,$km.
%, i.e. the travel distance perturbation is below the
%resolution limit. 
This effect is very small because in the shallow layer
$z<z_0$ the ray path is very close to vertical.

For the travel time reduction we find in an analoguous manner
\begin{eqnarray}
  \Delta\tau  && \equiv\tau_{\mbox{\scri\rm quiet sun}}
     -\tau_{\mbox{\scri\rm  plage}}=   
 2^{-3/2} f_1 m^{-1/2}{\alpha}^{-2}\,\frac{B^2/B_e^2}{1+B^2/B_e^2}\,\times \\
   &&
    \left\{ \,{e^{-\alpha\,z_t}} [2\alpha-a^{-1}(1+\alpha\,z_t)]
    -\,{e^{-\alpha\,z_0}}[2\alpha-a^{-1}(1+\alpha\,z_0)] 
    \right\} . \nonumber 
\end{eqnarray}

\begin{figure}
\centerline{
\includegraphics*[width=\columnwidth]{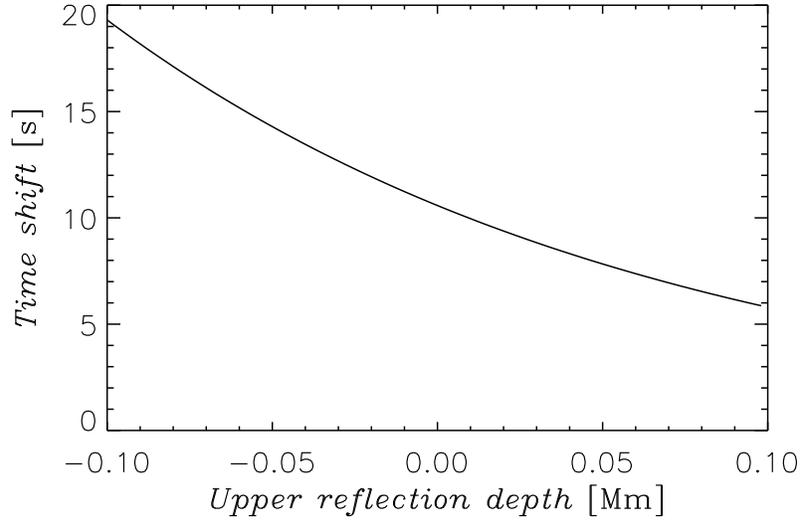}
}
\caption{Travel time reduction  
%[for what skip distance, or equivalently for what phase speed? !mjt] 
in seconds for $B=400\,$G as a function of  the depth of the upper reflection, 
in the limit of large skip distance, $d\gg z_0\simeq 1\,$Mm.}
\label{fig:travt_d}
\end{figure}

\begin{figure}
\centerline{
\includegraphics*[width=\columnwidth]{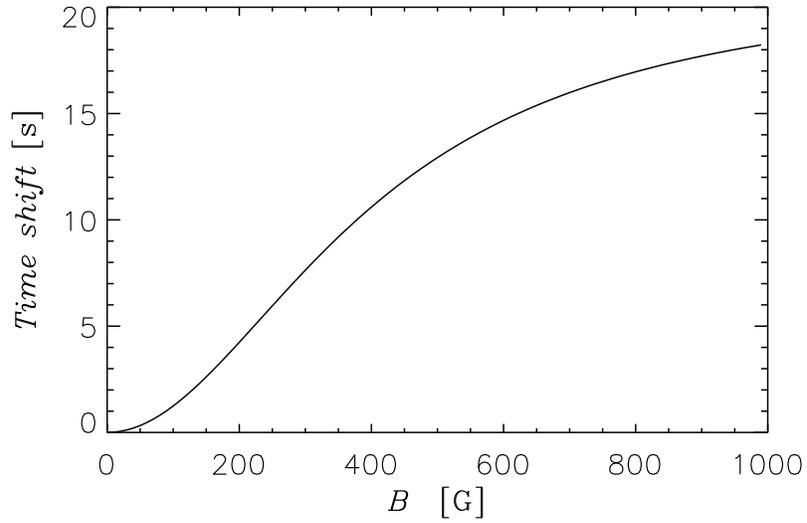}
}
\caption{Travel time reduction 
for upper reflection at unit optical depth, as a function of field strength,
in the limit of large skip distance, $d\gg z_0\simeq 1\,$Mm.}
\label{fig:travt_B}
\end{figure}

Figures \ref{fig:travt_d} and \ref{fig:travt_B} show $\Delta\tau$ as a function
of upper reflection depth $z_t$ and plage field strength $B$. The time shift
shows only negligible dependence on the travel distance, as a simple
consequence of the abnormal granulation layer depth $z_0$ being small compared
to the lower reflection depth $a$, so that the rays propagate nearly vertically
in the affected region. This, in turn, implies that the {\it relative} travel
time reduction decreases with travel distance, similar to the results of
\citet{Bruggen+Spruit}.  The overall magnitude of the travel time perturbation
is of the order of 10 seconds.  While this falls short of explaining the
observed shifts of 20--60 seconds (\citen{Chou:acoustic.imaging}), the result
shows that the sound speed perturbation introduced by abnormal granulation can
explain a  significant fraction of the measured time shifts.

The location of the upper turning point of the waves is a strong  function of
frequency, because of the variation of the acoustic cutoff frequency in the
lower photosphere. We find that the time shift shows a marked frequency
dependence (up to a factor 2 to 3 in the range 3 to 5 mHz), the travel time
reduction being larger for higher frequencies.  This is because higher
frequency waves are reflected higher up in the photosphere and so spend a
longer time in the layers affected by abnormal granulation.

As the suppression of turbulence, and, consequently, the perturbation of the
sound speed, depends strongly on the magnetic energy density, we expect that the
resulting time shifts should show a marked dependence on the value of the
assumed plage field strength. This is borne out in Figure~\ref{fig:travt_B}.

\subsection{Phase shift vs. envelope shift}

As we have mentioned in Section~2.1 above, the ray approximation breaks down
near the upper reflection point, where $\omega_c$ is not negligible. The
nearly
vertically propagating wave is then described by the dispersion relation (1),
resulting in different values for the group velocity $c_g$ and the phase
velocity $c_p$. From the dispersion relation it follows that

\begin{equation} c_p c_g=c^2  , \end{equation}
\begin{equation} c_g/c_p=1-(\omega_c/\omega)^2 . \end{equation}

Thus, the group velocity is less than the phase velocity, so that the wave
train spends a longer time in the abnormal granulation layer than each wave
crest does. This may be expected to give rise to a larger envelope time shift
compared to the phase shift. Quantitatively, the effect is rather sensitive to
the exact profile of $\omega_c$, but a numerical estimate using the model of
\citet{Guenther} shows that the resulting difference between phase and envelope
time shifts may reach a factor of 2 to 3, which is comparable to the findings
of \citet{Chou:acoustic.imaging}.

\subsection{Frequency shifts of global modes}

The upshot of the above results is that magnetic inhibition of turbulence in
plage areas has the effect of increasing the effective sound speed in a shallow
layer of depth $z_0\sim 1\,$Mm. This, in turn, leads to a reduction
$\Delta\tau$ in sound wave travel times. The dependence of $\Delta\tau$ on
magnetic field strength $B$ and frequency $\nu$ is in qualitative agreement
with the observations, but its amplitude is too low by a factor of 2 to 5. This
suggests that the effect considered here is just one, albeit non-negligible
contribution to the sound speed perturbation in plage areas. However,
additional effects such as those considered by
\citet{Zweibel+Bogdan:sound.fibrils} should also be limited to a shallow layer,
producing the same qualitative dependence.

A possibility to test these conclusions independently of the travel time
measurements is offered by the determination of global mode frequency shifts in
active regions from ring diagram analysis (\citen{Rajaguru+:ringdiag}).
Considering these shifts in the framework of our approach essentially
corresponds to inverting the problem: instead of fixing $k_x$ and $\omega$ and
looking for the skip distance $d$ and the travel time, we now fix $k_x$ and
$d$, and look for $\omega$. Equating the wavelength to the skip distance we
have $d=2\pi/k_x$; substituting this into equations (\ref{eq:wavepath}) and
introducing $\nu=\omega/2\pi$ we have $\nu=(m/a)^{1/2}/\pi=1/\tau$, as expected.
From this
\begin{equation} 
\Delta\nu/\nu=\Delta\tau/\tau=\nu\Delta\tau       \label{eq:dnunu}
\end{equation} 

As we have seen, $\Delta\tau$ increases by a factor of 2 or so in the frequency
range 3 to 5 mHz. Then equation (\ref{eq:dnunu}) shows that $\Delta\nu/\nu$
increases with $\nu$ faster than linear (approximately quadratically). This is
indeed confirmed by Figure~1 of \citet{Rajaguru+:ringdiag} in the given
frequency range. On the other hand, for frequencies below about 2.5 mHz the
observed behavior changes to a growth slower than linear. This is most likely
due to the non-negligible effect of spherical geometry on these low degree
modes.

According to equation (\ref{eq:dnunu}), $\Delta\nu/\nu$ should show the same
kind of dependence on the magnetic field strength as $\Delta\tau$,  cf.\
Figure~\ref{fig:travt_B}. This is indeed confirmed by Figure~2 of 
\citet{Rajaguru+:ringdiag}. (Note the different horizontal range on these
two plots.)

The dependence of the relative frequency shifts on $\nu$ and $B$ is, then,
qualitatively quite like what is observed. On the other hand, comparison of our
Figure~5 with Figures~1 and 2 of Rajaguru et al. shows that the observed amplitude
of the shift exceeds the predicted amplitude by about a factor of 2 for
$B=100\,$G. This is in accordance to the case of the travel time corrections
discussed above, and suggests that another effect of similar depth dependence
(most likely the mechanism proposed by \citen{Zweibel+Bogdan:sound.fibrils}) is
an important contributor to the sound speed perturbation in active regions.

\section{Discussion}

In this paper we have considered the effect that the magnetic suppression of
turbulence in plage areas should have on the propagation of acoustic waves. It
has been found that any such effect is confined to a shallow surface layer of
thickness below about 1\,Mm, where sound waves propagate approximately
vertically. Accordingly, the travel path of wave packets is  essentially not
modified by the effect considered. On the other hand, the weaker turbulence
implies higher sound speeds, leading to shorter travel times. This time shift
$\Delta\tau$ (i.e. the reduction of the travel time) is independent of the
travel distance, while it shows a characteristic dependence on the assumed
plage field strength, plotted in Figure \ref{fig:travt_B}. As a consequence of
the variation of the acoustic cutoff with height, and hence of the location of
the upper turning point, $\Delta\tau$ is expected to be significantly higher
for  higher frequency waves within the observed regime of 3--5 mHz. The lower
group velocity near the upper reflection point further leads to an increased
envelope time shift, as compared to the phase shift. $p$-mode frequencies in
plage areas are expected to be higher by a corresponding amount, 
$\Delta\nu/\nu=-\nu\Delta\tau$.

These characteristics of the time and frequency shifts are in accordance with
measurements, as reported e.g. by \citet{Chou:acoustic.imaging} and
\citet{Rajaguru+:ringdiag}. However, the overall calculated amplitude of about
ten seconds falls short of explaining the full observed time shifts, that
exceed one minute in the case of envelope shift. 

This is hardly surprising, as it is clear that magnetic suppression of
turbulence is just one of a number of effects at play in plage areas that may
influence sound propagation.  Another prime candidate for  explaining the time
shifts is the direct effect of the fibril magnetic field on wave propagation,
as considered by \citet{Zweibel+Bogdan:sound.fibrils}. Nevertheless, the
overall qualitative agreement between our model and the observations suggests
that any further important contribution to the time shifts should also be
confined to a rather shallow surface layer.

The validity of our approach is certainly constrained by the simplifying
assumptions made (ray approximation, frozen turbulence, isotropy). In
particular, the effect of turbulent motions on acoustic waves has been
considered by several authors (\citen{Stix:convected_waves};
\citen{Murawski:Budapest} and references therein) without making use of the ray
approximation. These promising approches, however, have not yet addressed the
problem of the effect of a magnetic suppression of turbulence on wave
propagation, studied in this paper. These more sophisticated models involve a
larger number of unknown parameters and functions, such as degree of anisotropy
or turbulence spectra. The effect of magnetic fields on these parameters is very
poorly known, so their introduction would lead to a high degree of uncertainty
into the problem at this stage.

Beside the direct effect of random velocity and magnetic fields on sound speed
below the surface, a more indirect effect from {\it atmospheric} flows and
magnetic fields can also be expected, as these effects change the boundary
conditions for wave reflection.

\citet{Cunha} suggested that acoustic wave scattering  from a nearby
sunspot could alone explain the apparent perturbed travel times found in plage
areas, even in the absence of any magnetic field outside the spot. This effect
could be separated from real physical effects, such as the one considered in
the present paper, by a careful study of the systematic dependence of any time
shift on the plage field strength. Exploratory work in this direction has been
made by \citet{Hindman} and \citet{Rajaguru+:ringdiag}. Further systematic
studies of this type will be needed until the relative contribution of
different effects to the observed travel time shifts can be clarified.

\acknowledgements

This research was partly funded by the UK Particle Physics \& Astronomy
Research Council under grant no.~PP/X501812/1 and by the Royal Society
International Joint Project RS15599. RE is grateful to M. K\'eray for patient
encouragement. K.P. acknowledges support from the ESMN network supported by the
European  Commission. R.E. and K.P. acknowledge support from the Hungarian
National Science Research Fund under grant no.~T043741. 

\section*{Appendix: The effect of a mean downflow}

Contrary to what was assumed throughout this paper, turbulence in the solar
convective zone is not homogeneous and isotropic. One potentially important
consequence of this on sound wave propagation is that, owing to a nonvanishing
correlation between velocity and density fluctuations (downflows are cooler and
denser), a mean net downflow will arise, i.e.\ $\langle v\rangle\neq 0$ even
though $\langle\rho v\rangle=0$. Numerical simulations indicate that
immediately below unit optical depth the Mach number of this downflow may reach
values  of $M\sim 0.1$. As such a mean flow will advect the propagating wave,
this might suggest to correct the r.h.s.\ of equation (\ref{eq:beta}) to $1\pm
M+\ov{\beta^2}$, the sign of $M$ depending on whether the flow is parallel or
antiparallel to the wave propagation. This may seem like a potentially
important correction.

However, it should be taken into account that between two skips the wave front
will first pass the shallow layers downwards (along the mean downflow), then
upwards (against the mean downflow), so the net effect cancels to first order
in $M$. For simplicity let us assume that the effective sound speed $C$, its
magnetic perturbation $\Delta C$ and the mean flow speed $V$ are constant in
the uppermost layer of thickness $z_0$ where the magnetic field is dynamically
important. Then the travel time reduction relative to the nonmagnetic case
is
\begin{eqnarray}
\Delta\tau && = \frac{z_0}{C+V}-\frac{z_0}{C+\Delta C+V}+\frac{z_0}{C-V}-
  \frac{z_0}{C+\Delta C-V}  \nonumber \\
  && = z_0\Delta C \left[\frac 1{(C+V)^2}
    +\frac 1{(C-V)^2}\right] + {\cal O}[(\Delta C/C)^2]
\label{eq:meanflow}
\end{eqnarray}

In the case with no mean flow ($V=0$) clearly $\Delta\tau=\Delta\tau_0\equiv 
2z_0\Delta C/C^2$. With this, series expansion of (\ref{eq:meanflow}) yields
\begin{equation}
\Delta\tau/\Delta\tau_0 = 1+3 M^2 +{\cal O}(M^3)
\end{equation}
For $M\equiv V/C\sim 0.1$ (a very generous upper limit, as in reality $M$
decreases very fast with depth) this shows that the reduction of the travel
time is increased by a few percents only compared to its value for homogeneous
turbulence.

%\bibliographystyle{solphys}
%\bibliography{abnormal}

\begin{ao}
\\
K. Petrovay\\
E\"otv\"os University, Dept.~of Astronomy\\
Budapest, Pf.~32, H-1518 Hungary\\
E-mail: K.Petrovay@astro.elte.hu
\end{ao}                   

\end{article}
\end{document}